\makeatletter
\disable@package@load{program}{}
\makeatother

\documentclass[pdflatex,sn-basic]{sn-jnl}

\usepackage{orcidlink}

\jyear{2021}%

\usepackage{lineno}

\usepackage{xcolor}

\theoremstyle{thmstyleone}%
\theoremstyle{thmstyletwo}%

\theoremstyle{thmstylethree}%

\begin{document}

\title[Atmospheric Science Questions for a Uranian Probe]{Atmospheric Science Questions for a Uranian Probe}


\author*[1]{\fnm{Emma K.} \sur{Dahl}\,\orcidlink{0000-0003-2985-1514}}\email{emma.k.dahl@jpl.nasa.gov}

\author[2]{\fnm{Naomi} \sur{Rowe-Gurney}\,\orcidlink{0000-0001-8692-5538}}\email{nrowe-gurney@ras.ac.uk}

\author[1]{\fnm{Glenn S.} \sur{Orton}\,\orcidlink{0000-0001-7871-2823}}\email{glenn.s.orton@jpl.nasa.gov}

\author[3]{\fnm{Shawn R.} \sur{Brueshaber}\,\orcidlink{0000-0002-3669-0539}} \email{srbruesh@mtu.com}

\author[4]{\fnm{Richard G.} \sur{Cosentino}\,\orcidlink{0000-0003-3047-615X}}\email{rcosenti@stsci.edu}

\author[5]{\fnm{Csaba} \sur{Palotai}\,\orcidlink{0000-0001-6111-224X}}\email{cpalotai@fit.edu}

\author[6]{\fnm{Ramanakumar} \sur{Sankar}\,\orcidlink{0000-0002-6794-7587}}\email{rsankar@umn.edu}

\author[7]{\fnm{Kunio M.} \sur{Sayanagi}\,\orcidlink{0000-0001-8729-0992}}\email{kunio.m.sayanagi@nasa.gov}

\affil*[1]{\orgname{Jet Propulsion Laboratory/California Institute of Technology}, \orgaddress{\street{4800 Oak Grove Dr}, \city{Pasadena}, \postcode{91109}, \state{CA}, \country{USA}}}




\affil[2]{\orgdiv{Royal Astronomical Society}, \orgaddress{\street{Burlington House, Piccadilly}, \city{London}, \postcode{W1J 0BQ}, \country{United Kingdom}}}

\affil[3]{\orgdiv{Department of Mechanical Engineering-Engineering Mechanics}, \orgname{Michigan Technological University}, \orgaddress{\street{1400 Townsend Drive}, \city{Houghton}, \postcode{49931}, \state{MI}, \country{USA}}}

\affil[4]{\orgname{Space Telescope Science Institute}, \orgaddress{\street{3700 San Martin Dr}, \city{Baltimore}, \postcode{21218}, \state{MD}, \country{USA}}}

\affil[5]{\orgdiv{Department of Aerospace, Physics \& Space Sciences}, \orgname{Florida Institute of Technology}, \orgaddress{\street{150 W University Blvd}, \city{Melbourne}, \postcode{32901}, \state{FL}, \country{USA}}}

\affil[6]{\orgdiv{Center for Integrated Planetary Science}, \orgname{University of California, Berkeley}, \orgaddress{\street{University Ave and Oxford St}, \city{Berkeley}, \postcode{94720}, \state{CA}, \country{USA}}}

\affil[7]{\orgname{NASA Langley Research Center}, \orgaddress{\street{1 Nasa Dr}, \city{Hampton}, \postcode{23666}, \state{VA}, \country{USA}}}


\abstract{The Ice Giants represent a unique and relatively poorly characterized class of planets that have been largely unexplored since the brief Voyager 2 flyby in the late 1980’s. Uranus is particularly enigmatic, due to its extreme axial tilt, offset magnetic field, apparent low heat budget, mysteriously cool stratosphere and warm thermosphere, as well as a lack of well-defined, long-lived storm systems and distinct atmospheric features. All these characteristics make Uranus a scientifically intriguing target, particularly for missions able to complete \textit{in situ} measurements. The 2023-2032 Decadal Strategy for Planetary Science and Astrobiology prioritized a flagship orbiter and probe to explore Uranus with the intent to “...transform our knowledge of Ice Giants in general and the Uranian system in particular” \citep{owl}. In support of this recommendation, we present community-supported science questions, key measurements, and a suggested instrument suite that focuses on the exploration and characterization of the Uranian atmosphere by an \textit{in situ} probe. The scope of these science questions encompasses the origin, evolution, and current processes that shape the Uranian atmosphere, and in turn the Uranian system overall. Addressing these questions will inform vital new insights about Uranus, Ice Giants and Gas Giants in general, the large population of Neptune-sized exoplanets, and the Solar System as a whole.}


\begin{center}
    \small \textit{Accepted for publication in Space Science Reviews}
\end{center}

\begingroup
\let\newpage\relax
\maketitle
\endgroup

\maketitle

\section{Introduction}\label{sec1}

The Ice Giants, Uranus and Neptune, have been visited just once by a dedicated spacecraft, in 1986 and 1989, respectively. Voyager 2 observed Uranus while the planet was at solstice and revealed an enigmatic world with unique characteristics, including but not limited to an apparent lack of internal heat release \citep{pearl_1990}, a warmer-than-expected thermosphere \citep{lindal_1987}, and a complex magnetic field offset and tilted from the planet's axis of rotation \citep{ness_1986}. In all, Uranus was revealed to be an extreme environment and a rich laboratory for testing theories of magnetic field interactions, atmospheric dynamics, planetary formation, and more. However, we have yet to revisit Uranus with a dedicated spacecraft equipped with a suite of instruments tailored to exploring this strange planet. In particular, Uranus's atmosphere presents a particularly scientifically rich area of the planet to explore and holds the key to addressing several major outstanding questions regarding the nature of the interior, interactions with the magnetic field, and formation and dynamical mechanisms that produce the planet that we see today. While remote-sensing instruments onboard a Uranus-orbiting spacecraft can and will generate data critical for addressing open questions regarding Uranus's atmosphere, an atmospheric probe will greatly enhance and further contextualize those observations and, in some cases, provide unique measurements that are only possible \textit{in situ}. 

Citing Uranus's suite of mysterious properties, technological readiness, and the great deal of community support, the 2023-2032 Oceans, Worlds, and Life: A Decadal Strategy for Planetary Science and Astrobiology (hereafter referred to as the 2023-2032 decadal survey), as compiled by the National Academies of Sciences, Engineering, and Medicine (NASEM), recommended a Uranus orbiter and probe (UOP) as NASA's top priority for flagship mission development over the next decade \citep{owl}. In concert with an orbiting spacecraft, the recommended probe will greatly enhance and increase the scientific return of the UOP mission, particularly through its direct measurements of Uranus's atmosphere. Such a mission is relevant to several of the priority science question topics and themes as outlined by the 2023-2032 decadal survey \citep{owl}:
\begin{itemize}
    \item \textit{Evolution of the protoplanetary disk.} What were the initial conditions in the solar system? What processes led to the production of planetary building blocks, and what was the nature and evolution of these materials? 
    \item \textit{Accretion in the outer solar system.} How and when did the giant planets and their satellite systems originate, and did their orbits migrate early in their history? How and when did dwarf planets and cometary bodies orbiting beyond the giant planets form, and how were they affected by the early evolution of the solar system? 
    \item \textit{Giant planet structure and evolution.} What processes influence the structure, evolution, and dynamics of giant planet interiors, atmospheres, and magnetospheres? 
    \item \textit{Exoplanets.} What does our planetary system and its circumplanetary systems of satellites and rings reveal about exoplanetary systems, and what can circumstellar disks and exoplanetary systems teach us about the solar system? 
\end{itemize}

To inform NASEM on the issues most important to the field, the planetary science community produced hundreds of white papers voicing support for various studies, instruments, missions, and emphasizing the most important questions to be addressed at various Solar System bodies. Within the new context of the highly-prioritized orbiter and atmospheric probe, we here highlight and reframe several community-sourced science questions expressed by three white papers that focused on the exploration of Uranus's atmosphere, and how these newly recontextualized goals can be addressed and enhanced by an \textit{in situ} probe. These white papers are ``Ice Giant Atmospheric Science" \cite{dahl_2020}, ``In Situ Probes in the Atmospheres of the Ice Giants" \citep{orton_2020_1}, and ``Probes in the Atmospheres of the Ice Giants" \citep{orton_2020_2}. In particular, these questions seek to address some of the biggest outstanding questions regarding Uranus's formation, the evolution of its atmosphere, and how atmospheric processes give rise to the planet we see today:

\begin{enumerate}
    \item What does Uranus’s atmospheric composition reveal about its migration and formation history? How can those measurements inform our understanding of the origin of the solar system and of Ice Giants in general? 
    \item How has the atmosphere of Uranus regulated its long-term thermal evolution? Why does Uranus appear to exhibit negligible internal heat release? 
    \item What is the role of moist convection in vertical heat transport in the Uranian atmosphere?
    \item What drives the long- and short-term chemical and photochemical processes that affect Uranus’s atmospheric composition and temperature profile at the top of the atmosphere? Do these processes influence the thermal evolution of Uranus’s thermosphere?
    \item How are meridional and zonal circulation patterns coupled, and how do they transport material and energy? How are these patterns of circulation maintained? 
    \item How does periodic seasonal forcing affect the state of the Ice Giant atmospheres, especially in the case of Uranus’s extreme axial tilt? 
\end{enumerate}

Section \ref{sec2} contains the discussion of each community-supported science question, which includes the scientific motivation, key measurements that are necessary and/or helpful for addressing the question, any resolution requirements for those measurements. Occasionally, if a remote sensing measurement might be required or would complement the \textit{in situ} measurements, this is highlighted as well. Section \ref{sec3} expands on the instruments required to obtain the measurements required to address each science question.
Section \ref{sec4} includes a top-level summary of the recommended instrument suite optimized for addressing these atmospheric science questions, miscellaneous requirements for entry location, measurement cadence, etc., and a review of the big-picture insights that can be provided by an atmospheric probe outfitted with such an array of instruments.

\section{Questions, Measurement Requirements, and Suggested Instruments}\label{sec2}

\subsection{Question 1: Composition and Formation History}\label{sec_q1}

\subsubsection{Background and Motivation}

The mechanisms that formed the Solar System, and the Ice Giants in particular, are active areas of study (e.g., \citet{atreya_2020}). Competing theories of planetary formation mechanisms fall into four main categories: the currently more generally-favored model of core accretion \citep{pollack_1996,Hubickyj_2005}, the theory of disk instability \citep{boss_1997,boss_2002}, the photoevaporation model \citep{guillot_2006}, and the CO snowline model \citep{alidib_2014}. Since the composition and temperature of the protosolar nebula was highly radially and temporally dependent, the abundances of certain diagnostic species and their isotopes relative to solar, cometary, or other planetary abundances can greatly aid in constraining the conditions of Uranus's formation. Especially important are the abundances of noble gases, which can inform the nature of the formation process and whether it occurred in a step-wise fashion or more instantaneously and can only be measured \textit{in situ}.

The process of core accretion first necessitates the condensation of solid and/or icy planetesimals and the subsequent, rapid infall of large amounts of gas onto that solid core once it becomes massive enough. The possibility of this mechanism was originally limited by the length of its predicted time frame, which was at or longer than the lifetime of the protosolar nebula. However, more recent models have found that this process could take place on timescales short enough to render it possible and even likely \citep{rice_2022}. Depending on the way in which volatiles were incorporated into the solids that eventually feed condensing planitesimals, different compositional signatures will be reflected in the planet's modern bulk abundances \citep{mousis_2020}. Under the assumption that there is no process leading to differences in fractionation between volatiles, the theory of core accretion via amorphous ice predicts homogeneous enrichments of O, C, N, S, Ar, Kr and Xe since their trapping efficiencies within amorphous ice are all very similar \citep{owen_1999, barnun_2007}. In the case of core accretion by way of clathrates, a strong variation of trapping efficiencies between species is implied \citep{mousis_2010}, resulting in elemental abundances that vary as a function of both trapping temperature and dependence upon the availability of crystalline water, which is necessary to achieve either full or partial clathration \citep{gautier_2001,mousis_2014,mousis_2012}.

The disk instability model of planet formation predicts that instead of a more gradual condensation of solids and subsequent rapid collapse of gas, a gravitational instability within the protosolar disk causes a self-gravitating collapse of material into a planet much more quickly, on the order of $\sim$1000 years \citep{boss_2002}. When applied to our Solar System, this theory predicts enrichments of O, C, N, S, Ar, Kr and Xe relative to protosolar abundances due to two main factors: the settling of dust grains prior to mass loss \citep{mousis_2018}, and the possibility of photoevaporation of the gaseous envelopes of any young planet beyond 5-10 AU by nearby, newly-formed OB stars \citep{boss_2002_2}. While the theory of disk instability cannot yet be ruled out as \textit{the} or one of the mechanism(s) of formation for Uranus and the gas giant planets in general, there are several discrepancies within models implementing this theory that are difficult to account for \citep{mousis_2020}. 

The photoevaporation model states that after forming at lower temperatures farther from the Sun, ice grains would have adsorbed Ar, Kr, and Xe; subsequently migrating inward, these grains encountered increasing temperatures and those captured noble gases would have been released in this new location. These grains would essentially transport these noble gases to the regions where the Ice Giants might have formed, allowing a young Uranus to accrue supersolar noble gas enhancements, albeit smaller enrichments than those that would originate from solids containing O, C, N, and S alone \citep{guillot_2006,mousis_2020}. 

Lastly, the CO snowline (around 30 AU, where the temperature of the protosolar nebula was equivalent to CO's freezing temperature) model predicts that both Uranus and Neptune were formed past the CO snowline both by solids containing C and O enrichments and gas depleted of nitrogen. Under these circumstances, Uranus's atmosphere would contain equally depleted levels of Ar and N, but supersolar levels of Kr, He, S, and P. In comparison, C and O abundances would be very high/. Notably, this scenario would produce a D/H ratio that would be low enough to reflect observed values (e.g. \citet{Feuchtgruber_2013}) which, along with a higher O abundance, is often difficult for other models to produce \citep{alidib_2014}.

Understanding how the planets of our Solar System formed in general is a major outstanding question, but the formation history of the Ice Giants and Uranus in particular hold the key to disentangling the exact process that resulted in the current distribution of Gas Giants, in terms of size, distance from the Sun, and bulk compositions. Due to their larger distance from the Sun than Jupiter and Saturn, Uranus and Neptune had longer available formation timescales, and so might have undergone different processes under different conditions. Additionally, understanding how Neptune-sized planets form in general is not only necessary to understand the evolution of our own Solar System, but also exoplanetary systems where Neptune-sized exoplanets are plentiful \citep{Batalha_2013}. In order to constrain models attempting to balance Uranus's total hydrogen/helium mass, their metallic composition, and the time frame of formation, the bulk abundances of several significant species must be measured accurately.

\subsubsection{Measurement Requirements}

To address this question and differentiate between these formation and migration theories, it is necessary to determine the bulk abundances of a series of elements and isotopes \citep{mandt_2020}. In general, the required degree of accuracy stems from the need to compare these measurements to other known quantities throughout the Solar System. In particular, comparisons to known protosolar abundances will allow for discrimination between formation models that estimate certain bulk compositions. 

Noble gases and their isotopic ratios, namely He, Ne, Ar, Kr, and Xe, are measurable only with an \textit{in situ} probe. An exception is He, in that it can be partially derived from far-IR observations if temperature and the para-H$_2$ abundances are also constrained \citep{fletcher_2020}, albeit at a much lower level of accuracy than can be obtained \textit{in situ}. Abundances of these noble gases relative to H, along with their isotopes, will aid in differentiating between the aforementioned planetary formation theories by enabling the identification of the subresevoirs of material in the protosolar nebula that fed planet formation. The He measurement should be accurate to at least $\pm$2\% to enable a comparison to those made by the Galileo probe and to levels in Jupiter's atmosphere. Similarly, Ne, Xe, Kr, and Ar should be measured at $\pm$1\% to match the accuracy of those values known for the rest of the Solar System \citep{orton_2020_2}.

Measurements of the bulk abundances of C, N, S, O, and P down to a pressure of at least 10 bars will inform both the nature of the objects that condensed to form the planet and the formation location relative to the carbon monoxide ice line. These elements should be obtained to accuracies of 10\% or better, which is close to the current uncertainty of their protosolar abundances \citep{mousis_2018, atkinson_2020, pre_decadal_survey, orton_2020_2}. Some of these bulk abundances can be obtained through tropospheric measurements of CO and PH$_3$ (to levels of $\pm$5\% for comparison to known values \citep{fletcher_2009,mousis_2014}), which together can help constrain the deep H$_2$O profile, since a probe could not reach the depths below the water condensation level. \citet{Visscher_2005} were able to use a combination of tropospheric PH$_3$ and CO measurements to produce upper and lower limits to the deep H$_2$O/H$_2$ abundance at Saturn through the application of thermochemical equilibrium and kinetic calculations. Similarly, \citet{cavalie_2020} provided better constraints on the deep abundance of H$_2$O at Uranus. They reproduced CO observations with a thermochemical and diffusion model that accounted for an inhibition of convection brought on by the mean molecular mass gradient resulting from H$_2$O condensation at depth. 
To determine the nature of the reservoir of material within the  protosolar nebula where Uranus was formed, the isotopic ratios of these species should also be obtained at certain accuracy levels, namely D/H and $^{15}$N/$^{14}$N at $\pm$5\% \citep{mousis_2016}, $^{3}$He/$^{4}$He at $\pm$3\% (at least as good as measurements made by the Galileo neutral mass spectrometer \citep{mahaffy_1998}), and $\pm$1\% for $^{17}$O/$^{16}$O, $^{18}$O/$^{16}$O, and $^{13}$C/$^{12}$C to enable comparisons to known Solar System values for those isotopes \citep{mousis_2018, atkinson_2020, orton_2020_1, orton_2020_2} 

The distinction between various formation models and their implications for relative abundances that would be measured by an in-situ probe was elucidated most recently by \citet{mousis_2022}. The clearest distinctions between different formation models lie in measurements of different enhancements or depletions of specific elements or families of elements with respect to their expected abundances in the protosolar nebula. Although schematic, Figure 2 of \citet{mousis_2022} indicates enhancements and depletions resulting from the different models to be around factors of several. Therefore, the critical uncertainty controlling the differentiation between these models lies in the uncertainty of abundances in the protosolar nebula itself, which is on the order of 10\% \citep{lodders_2009}. Absolute measurements more accurate than 10\% would not provide additional information.
For Helium in particular, a case can be made for a measurements at the $\sim$2\% level that is commensurate with the uncertainty of the Galileo probe Helium Abundance Detector (HAD) \citep{von_zahn_1998}. It is only the HAD measurement that decisively noted that the He abundance was below what one expected in a protosolar nebula, and thus subject to dissolution in the deeper atmosphere.  Current interior models (e.g., \citet{guillot_2005, lambrechts_2014, helled_2020}) do not expect this to be the case for Uranus or Neptune, and so it is important to determine its abundance to the same level of uncertainty as for the Galileo mission.

\subsection{Question 2: Internal heat budget and thermal evolution}\label{sec:2.2}

\subsubsection{Background and Motivation}

Uranus's heat budget, or the balance between solar insolation and radiative heat loss, is enigmatic among Gas Giants as it appears to be half that of Neptune's despite being closer to the Sun and receiving more radiant solar energy \citep{pearl_1990, pearl_1991}. Uranus's atmosphere is the medium through which residual gravitational potential energy from planetary formation is released, and thus, can heavily influence the planet's heat budget and its thermal evolution. In general, the processes that transport this internal heat energy through gas giant atmospheres are cumulus convection and radiation \citep{li_2015}; however, some other process might also be occurring in Uranus's atmosphere that either inhibits this type of heat flow or causes atypical patterns of heat transport altogether. 

Several hypotheses have been put forward to explain Uranus's apparent low internal heat flux and luminosity. Some deep compositional gradient and resultant thermal boundary layer, possibly initiated by a giant impact early in Uranus's history \citep{kegerreis_2018}, might be responsible for trapping heat deep in the atmosphere and inhibiting its vertical transport \citep{nettelmann_2016, vazan_2020, scheibe_2021}. For example, the presence of even a thin conductive layer in Uranus's interior has been predicted to dramatically affect planetary cooling \citep{scheibe_2021}. Another possibility is simply the fact that various potential issues with the Voyager measurements of Uranus's luminosity misled estimates of the amount of internal heat. Possible issues include limitations on phase angle coverage, possible calibration issues for Voyager instruments \citep{li_2018}, and/or temporal changes in reflectance \citep{lockwood_2019} since those measurements were made. Or, we might simply be observing Uranus during a transient, quiet period of heat release relative to Neptune; a discussion of the meteorological causes of such a quiescent period can be found in Section \ref{sec:2.3.1}. \citet{Li_et_al_2015} found that Saturn's globally-averaged radiant power to space increased by $\sim$2\% during its 2010--2011 Great White Spot outbreak, which renders it possible that a quiescent and less-stormy Uranus might have appeared to have a low outgoing heat flux during the Voyager epoch.

\subsubsection{Measurement Requirements}\label{sec:2.2.2}

To test the above hypotheses, the temperature/pressure profile should be measured to identify any regions with sub- or super-adiabatic lapse rates. Identifying deviations from adiabatic lapse rates can reveal how and to what degree internal heat is being transported farther down in the atmosphere. To accurately measure these temperature/pressure profiles, pressure measured with an accuracy of $\pm$1\% and temperature to $\pm$1 K will be required \citep{orton_2020_1}. These precision requirements arise from the need to differentiate between different lapse rates and degrees of atmospheric stability as observed by Voyager 2 and predicted by works such as \citet{guillot_1995}, \citet{leconte_2017}, and \citet{cavalie_2020} in the middle and deep troposphere and \citet{orton_2014} and \citet{roman_2020} in the upper troposphere and stratosphere, as well as the need to locate any radiaitve boundary layers. Using radio occultation measurements, Voyager 2 measured temperature down to $\sim$2 bars with an accuracy of 1 K and at best 1\% accuracy for pressure \citep{lindal_1987}, and so a comparable accuracy is necessary to understand how Uranus's temperature profile has changed since 1986. A sampling frequency of 1 Hz will allow for multiple ($\sim$10, depending on the probe's descent rate) measurements per scale height (e.g., \citet{sayanagi_2020}).

Additionally, the vertical profile of condensable species abundances, especially those containing C, N, S, O, and P, particularly CH$_4$ and H$_2$S (H$_2$O is unlikely to be feasible), ideally at the levels described in Section \ref{sec:2.1.3}, but for these purposes 10-20\% will suffice since matching well-constrained Solar System or protosolar values is not required. Measuring condensable abundances multiple times per scale height is the minimum sampling frequency needed to address this science question and contextualize the measured lapse rate, but higher sampling frequencies are highly desirable to provide better constraints to extant model predictions \citep{ferri_2020}. Such profiles will also be valuable for identifying any potential sources or sinks of those species that might be populating the compositional/thermal boundary layer (e.g., \citet{cavalie_2020}). Additionally, a microwave radiometer onboard the orbiter would be better able to access pressures and reservoirs of gases that the probe cannot. 

Also necessary is the vertical profile of horizontal winds to $\pm$10 m/s, which can be obtained through Doppler measurements of the probe's position as it descends through the atmosphere \citep{sayanagi_2020}. These measurements will help reveal the way in which wind shears with depth, what sort of waves it might encounter, and will allow for linking wind shear and the horizontal temperature gradient through the thermal wind relationship \citep{orton_2020_2}.

The degree of vertical convection, and therefore the convective capability at cloud condensation pressures, is critical to understanding the way in which atmospheric stability affects heat transfer \citep{banfield_2005}. This information can be obtained through measurements of the speed of sound to accuracies of $\pm$1\% in a given region of the atmosphere, which can then can be used to derive the ortho-para H$_2$ fraction at that location \citep{orton_2020_1}. 

To identify the direction of heat flow in the atmosphere and any resulting differences in buoyancy, the net atmospheric radiative energy balance should be measured by finding the difference between downwelling solar energy and upwelling thermal infrared energy \citep{atkinson_2020}. In order to constrain those two contributors to the radiative energy flux, the altitude profile of thermal infrared light and the amount of absorbed visible sunlight should be determined. The amount of absorbed sunlight can be found once the planet's Bond albedo is characterized, which requires remote sensing measurements from the orbiter. Visible flux, in both directions, should be measured from 0.4--5 $\mu$m to fully cover the contribution from solar flux, and the IR flux should be measured over 4--50 $\mu$m, which is much wider than similar requirements for Neptune \citep{pre_decadal_survey} to account for Uranus's possibly much lower themral emissoin. Both fluxes should be acquired with a spectral resolution ($\lambda/\Delta \lambda$) of 0.1--100 over their respective wavelength ranges, a flux resolution of $\sim$0.5 W/m$^{-2}$, and an accuracy of $\pm$1\% \citep{pre_decadal_survey, orton_2020_1}.

\subsection{Question 3: Role of Moist Convection}

\subsubsection{Background and Motivation}\label{sec:2.3.1}

The dominant mechanism of vertical heat transport in the troposphere, aside from incoming solar radiation and outgoing thermal heat from the interior, is cumulus convection driven by the condensation of CH$_4$, H$_2$O, NH$_4$SH, and possibly NH$_3$ and/or H$_2$S \citep{hueso_2020}. This process likely plays an important role in the energy budget of the atmosphere and in storm generation \citep{sromovsky_2005}. Could this convection, and its associated release of latent heat and convective available potential energy (CAPE), be behind the episodic storms observed in Uranus’s atmosphere?

In giant planets’ hydrogen-dominated atmospheres, condensable species have a significantly higher molecular weight than the background H$_2$-He atmosphere. Consequently, the presence of heavier vapor made up of those condensables can suppress cumulus convection for prolonged periods and may allow for the build up of an enormous amount of convective available potential energy (CAPE) before convection is triggered \citep{li_2015}. This process is thought to be behind the roughly 30-year cycle on Saturn's Great White Spot outbreaks on Saturn \citep{sanchez-lavega_2018, li_2015}. When sufficient CAPE is accumulated and latent heat release becomes capable of overcoming the stabilizing heavy vapor, an episodic cumulus storm may erupt, enhancing the heat flux to space. On Uranus, the interval between such episodic storms may be decades or centuries. See Section \ref{sec:2.6.1} for a discussion of  periodic and potentially seasonal cloud features that have been observed on Uranus.

During a quiescent period, storm activity may be very low and the resulting heat transport to space may fall to levels comparable to the incoming solar flux.
Alternatively, or perhaps in conjunction with convective suppression from stabilizing layers of high vapor abundances, more deeply seated and less efficient radiative heat-transporting layers may exist on Uranus. The effect of such a quiescent period may be hiding a warmer-than-expected interior, while simultaneously appearing as a sluggish and bland upper troposphere above a much deeper and more convectively active layer. For more details on moist convection in gas giant atmospheres and associated phenomena, see \citet{palotai_2022}.

Uranus and Neptune appear to harbor large reservoirs of condensable species (e.g., CH$_4$, H$_2$O). When these vapors condense in an upwelling air parcel, they release latent heat, with H$_2$O being particularly energetic. However, water clouds are situated deep in the troposphere, perhaps at 50--200 bar, depending on the global abundance of O \citep{cavalie_2020}. CH$_4$ clouds condense much higher, around 1.3 bar \citep{irwin_2022}, and can be easily sampled by an entry probe assuming the probe passes through these ephemeral and localized clouds. The purported globally distributed H$_2$S cloud deck around 3--6-bar will also be accessible to the entry probe; most probe designs target at least the 10 bar pressure level \citep{mousis_2014, mousis_2018, orton_2020_1, Simon_et_al_2020}. Thermochemical equilibrium models suggest that a deeper NH$_4$SH cloud layer exists around 30--40 bar \citep{Weidenschilling_and_Lewis_1973, Atreya_and_Wong_2005}, but this may be too deep for the probe to measure within the constraints of battery power, telecommunication restrictions, or increasing radio wavelength opacity at and below that pressure level due to the presence of NH$_3$. However, sampling the upper CH$_4$ and H$_2$S cloud decks may be sufficient to constrain some of the circulation hypotheses, aerosol properties, and will provide invaluable data on the zonal wind profile. 

Moist convection may also play a role in forming the dark anticyclonic spots, more commonly found on Neptune (e.g., \citet{smith_1989}), but these have also been more rarely observed on Uranus \citep{hammel_2009}. Alternatively, these dark spots may form as a result of baroclinic or barotropic instabilities. These vortices are usually associated with bright companion clouds that sometimes exhibit rapidly changing morphologies, perhaps linked to convective activity. These spots and their companion clouds are much more rare on Uranus, which suggests they are linked to a more vigorously convective period, such as may currently exist on Neptune. It is not yet known if their dark appearance (in blue wavelengths; nearly undetectable in red wavelengths) are a result of lower aerosol abundance (cloud-clearing) in the H$_2$S cloud deck, or a change in the physical character of the aerosols there \citep{hadland2020}. However a recent study using the Multi Unit Spectroscopic Explorer instrument (MUSE) of Neptune's NDS-2018 dark spot \citep{irwin_et_al_2023} ruled out the cloud-clearing hypothesis. It remains to be seen if Uranus's dark spots are also a result of different aerosol properties like Neptune's but, given their similar atmospheric compositions, it is a likely scenario. While it is extremely unlikely a probe will descend into a dark spot, measuring representative aerosol populations at the altitudes these dark spots are observed will help constrain radiative transfer models. 

\subsubsection{Measurement Requirements}\label{sec:2.3.2}

Establishing a vertical profile of the temperature and composition of the atmosphere as a function of probe altitude is mandatory to determine the stability of the atmosphere against convection. Temperature and pressure should be captured at accuracies of $\pm$1 K and $\pm$1\%, as previously stated. Condensables, such as CH$_4$, H$_2$S, and NH$_3$, should be measured to accuracies at approximately $\pm$20\%, down to at least 5 bars, and at least multiple times per scale height \citep{sayanagi_2020}. These measurements will establish what parts of the troposphere, if any, contain super-adiabatic lapse rates, which can only exist due to the convectively-suppressing molecular weight gradient effect from condensable vapors of a sufficient concentration (i.e., specific humidity) \citep{Lian_Showman_2010}. Additionally, measurements of the ortho-para H$_2$ abundance as a function of altitude can be used to quantify the degree of vertical mixing \citep{banfield_2005}. Determining vertical profiles of these parameters will also help determine the amount of convective potential in the atmosphere and quantify the atmospheric stability.

In a paper detailing the science case and design for a small probe to be carried along with a larger and more heavily equipped probe, \citet{sayanagi_2020} recommends an alternative variable accuracy for the temperature measurement: $\pm$5 K between 1 and 10 mb (stratosphere), $\pm$2 K between 10 and 100 mb (stratosphere to tropopause), and $\pm$0.1~K at greater pressures to meaningfully distinguish lapse rates. For an atmospheric descent that takes 2,235 seconds between 100 mbar and 10 bar, a sampling rate of 1 Hz for both temperature and pressure measurements will ensure that at least 10 measurements are performed every scale height.

Doppler wind measurements, through the use of ultrastable oscillators on the probe and carrier spacecraft, can in principle detect the presence of atmospheric waves \citep{seiff_1997}. A Doppler wind experiment will establish a vertical profile of the horizontal winds (e.g. \citet{atkinson_1996, atkinson_1997, atkinson_1998}). These winds and waves, measured to an accuracy of $\pm$10 m/s \citep{orton_2020_1}, provide clues to atmospheric dynamical processes occurring at the probe entry site and are important in determining a more accurate measurement of zonal jet speeds. Acceleration measurements suggested by \citet{sayanagi_2020} need to span a large range, from 0.01 m/s$^{2}$ for a reconstruction of horizontal winds of $\pm$10 m/s, and up to 300 G during entry. A sampling frequency at greater than 50 Hz is specified in their report.

\subsection{Question 4: Drivers of Thermal Flux in the Upper Atmosphere} \label{sec:question4}

\subsubsection{Background and Motivation}

Above the stratosphere, Uranus’s thermosphere was measured by Voyager 2 to be warmer than expected \citep{broadfoot_1986}, but has been cooling at a rate of 8 K/yr since 1997 \citep{melin_2019,melin_2020}. Possible explanations for the thermosphere's thermal evolution include magnetic effects, such as auroral heating and the effects of daily magnetic field reconnections (e.g. \citet{cao_2017}), or some sort of atmospheric process. 

Despite receiving lower levels of insolation than Jupiter and Saturn, Uranus has relatively active photochemical processes taking place in the upper atmosphere. In the stratosphere and above, lofted and UV-dissociated CH$_4$ molecules generate an abundance of complex hydrocarbons and photochemical products \citep{moses_2017,moses_2018,moses_2020}. These products can effectively absorb solar energy from above and inhibit the release of heat from deeper in the atmosphere. Uranus's CH$_4$ homopause is lower than Neptune's, at regions where diffusion timescales are large \citep{moses_2018}. Meridional stratospheric circulation is also an unlikely candidate for the cause of this uplifting, since it also appears to be too lethargic \citep{flasar_1987}. It is therefore likely that some more deeply-rooted mechanism might be lofting that CH$_4$ to levels where it can be affected by solar radiation. It is also possible that seasonal changes to the eddy diffusion coefficient, driven by rising temperatures, increased the amount of hydrocarbons in the upper atmosphere since Voyager visited \citep{herbert_1987, bishop_1990} and in turn increased that region's cooling efficiency, thereby allowing the thermosphere to cool \citep{melin_2020}.

Identifying the dynamical mechanisms that influence the cooling and radiative processes in the upper atmosphere is key to narrowing down the possible causes of the enigmatic thermal evolution of Uranus’s thermosphere. First, measuring and characterizing the sources, sinks, and behavior of the products of CH$_4$ photolysis is necessary in order to use them as dynamical tracers. Additionally, CO has been found in the stratosphere above the 100-mbar level \citep{cavalie_2014} and could be used as a dynamical tracer, as well as HCN if it is detected in Uranus's upper atmosphere, as it has been in Jupiter's \citep{moreno_2003,cavalie_2023}.

\subsubsection{Measurement Requirements}

A probe can help disentangle the magnetic and atmospheric influences on the thermal evolution of the upper atmosphere by complementing remote sensing measurements of the top of the atmosphere. In particular, a probe can explore deeper dynamical effects and measure species abundances below $\sim$100 mbar (the point at which the probe will be stable enough to make accurate measurements; e.g., \citet{sayanagi_2020}). Wind speeds and cloud motion, measured within $\pm$5 m/s to compare to dynamical models \citep{orton_2020_1}, will both be important characteristics to measure as the probe descends through the atmosphere, and useful for assessing the degree of circulation and uplift in regions below the stratosphere. Temperature/pressure profiles, with the same constraints as previously discussed, will be important for directly assessing the nature of heat transport in the region where the probe is dropped. The vertical composition profile of both elemental abundances and condensable species, especially CH$_4$, will be necessary to within $\pm$10\% \citep{orton_2020_1}; again, with as many measurements of these values per scale height as data volume allows. 

If possible at altitudes at and above the 100-mbar level, identifying the particle optical properties, size distributions, and number and mass densities, as well as the opacity, shapes, and composition of any photochemical hazes in the stratosphere will be critical to constrain the sources and sinks of the efficiently-radiating hydrocarbons \citep{moses_2020}. However, the probe will likely be unable to accomplish these measurements in the stratosphere due to the timing of parachute deployment. Therefore, the orbiter should be equipped to take similar or complementary measurements to characterize the photochemical products of CH$_4$ photolysis.

\subsection{Question 5: Zonal and Meridional Circulation Patterns}

\subsubsection{Background and Motivation}

The probe's suite of instruments will primarily sound Uranus's tropsophere. This region of the atmosphere harbors zonal wind patterns observed for both Ice Giants that are markedly different in three main ways from those observed on the Gas Giants: 1) There is a single broad equatorial jet with retrograde motion relative to the planet's rotation; 2) There is only one prograde mid-latitude jet per hemisphere; 3) The width of these Ice Giant jets span up to several tens of degrees in latitude with velocities up to $\sim$700 m/s, making them both wider and faster than jets found on Jupiter and Saturn \citep{sromovsky_2005, ingersoll_1990}. Uranus in particular appears to contain belts and zones of darker and brighter reflectivity (e.g. \citet{sromovsky_2012}), but these are not closely correlated with the position or latitudinal gradient of the jets, as is true for Jupiter (e.g., \citet{garcia_2001}). Lastly, the jets on the Ice Giants do not penetrate as deep into the interior as they do on the Gas Giants (1,000 km at the Ice Giants vs. 3,000 km at Jupiter and 9,000 km at Saturn \citep{Kaspi_et_al_2013, Kaspi_et_al_2020}). The differences between the jets of the Ice Giants and those of the Gas Giants, particularly on Uranus under the influence of its extreme seasonal forcing, the cause of those differences, and how they influence other processes of atmospheric circulation are open questions.

With fewer cloud features available for tracking, combined with fewer observations, and large distances from Earth, we do not have a rich spatial and temporal data set to compare to the closer and more frequently visited Gas Giants. From a dynamics perspective, additional questions arise about Uranus's jets: 1) Do they violate the Rayleigh-Kuo and Charney-Stern stability criteria by large margins as they appear to do on Jupiter? 2) What is the magnitude of the eddy flux momentum---which powers the jets on the Gas Giants---into or out of the jets \citep{Salyk_et_al_2006, Del_Genio_et_al_2007}? Is this flux driven by moist convection as is thought on Jupiter and Saturn \citep{Ingersoll_et_al_2000}? We must keep in mind that zonal wind measurements on Uranus do not even span a complete Uranian year, therefore 3) Does this flux vary greatly over time, and if so, how may that affect the zonal and meridional wind circulations?  

The lower number of jets on the Ice Giants compared to the Gas Giants ($\sim$16 vs. 10 hours) are likely produced by the combination of the smaller planetary radii and slower rotation rates. In short, the Rhines length, which is a quantity that provides an estimate of zonal wind width, is much smaller than the Gas Giants' planetary radii (resulting in more jets), but comparable to the Ice Giants' planetary radii \citep{cho_1996}. However, this does not explain why the equatorial jet is retrograde-directed, but may be a consequence of reduced convective activity near the equator, unlike that seen on the Gas Giants \citep{Lian_Showman_2010}. 

The zonal jets are expected to decay upwards into the upper troposphere and, at much deeper levels below the troposphere, downwards toward the interior, but neither the upper nor lower decay profile is well constrained \citep{conrath_1990, Kaspi_et_al_2013, fletcher_2022}. However, a recent ground-based observation of Neptune from the Atacama Large Millimeter/submillimeter Array (ALMA) \citep{carrion-gonzalez_2023}, found retrograde stratospheric jets in the range of 170 m/s to 190 m/s near the equator at pressures of 2 mbar and 0.4 mbar, respectively. These wind speeds are in agreement with stellar occultation measurements and expectations from the thermal wind equation. Given the similarity between Uranus's and Neptune's tropospheric zonal jets, we would expect to find similar stratospheric zonal jets at Uranus. Measuring vertical wind shear in the upper atmosphere will provide much needed altitude constraints for tracking features from remote sensing observations. Wind shear will also provide an indication of the horizontal temperature gradient via the thermal wind relationship \citep{Holton_2004}. A probe's Doppler wind experiment will be able to provide good constraints on the wind profile in the stratosphere and upper troposphere, although the probe will long cease functioning before reaching either the water cloud layer or the depth where zonal jets are expected to decay \citep{Kaspi_et_al_2013}.

Interestingly, Uranus's equatorial jet around equinox in 2012 revealed a braid-like wave feature that appears unique in the Solar System \citep{sromovsky_2012}. Slightly north of Jupiter's equator, cloud features are often observed that may be manifestations of an equatorial Rossby wave. We do not know what type of wave might explain Uranus's equatorial braid, but it did not survive for more than a few years. This feature, and our experience with the Galileo probe entering a hot spot on Jupiter devoid of water signatures, needs to be considered when studying Uranus's equatorial dynamics and data obtained from a probe entering this region. Furthermore, several wave-like phenomenon have been observed on Jupiter which provide information about dynamical stability of the region and the processes that drive atmospheric flows \citep{Simon2018}. Similarly, the Juno spacecraft observed a plethora of small scale wave-like phenomenon on Jupiter \citep{orton_2020_3}, providing details about local dynamics. Understanding the nature and distribution of such features on Uranus will enable key insights into both the local and global dynamical structure of the Uranian atmosphere, and formation of dynamical instabilities.

From efforts to reconcile observed temperatures, opacities, zonal wind profiles, and the results of numerical modeling, a new hypothesis has emerged to explain these varied characteristics in Jupiter's atmosphere: the presence of a double-stacked set of overturning cells in the troposphere \citep{Ingersoll_et_al_2000, Showman_dePater_2005, fletcher_2020, Yamazaki_et_al_2005, Zuchowski_et_al_2009}. In the upper troposphere ($\sim$0.1--1 bar), large-scale meridional circulation suggests upwelling occurs in the midlatitudes with downwelling occurring over the equator and poles. Deeper, a mid-troposphere cell (1 bar to an undetermined depth) exists and is characterized by counter-rotation to that of the upper tropospheric cell. Such stacked, counter-rotating cells have been invoked to explain seemingly conflicting results of observations in Jupiter's belts and zones. The Juno spacecraft has offered evidence to support the double-stacked hypothesis on Jupiter \citep{Fletcher_et_al_2021_Juno}. However, a dynamically consistent picture that connects aerosol and temperature observations together with the zonal-mean circulation is currently incomplete for the Ice Giants. An orbiter will be necessary to determine if the double-stacked cell hypothesis is valid at Uranus, with any degeneracy in solutions provided by remote sensing potentially being resolved with probe measurements.

\subsubsection{Measurement Requirements}

The measurement requirements to address the circulatory and zonal dynamics of Uranus's troposphere are identical to those in Section \ref{sec:2.3.2}. The one additional requirement is a measurement of ortho-para H$_2$ fraction via speed of sound measurements at accuracies of $\pm$1\%. In this case, the ortho-para H$_2$ fraction will act as a dynamical tracer in the upper troposphere. The ortho-para H$_2$ ratio is higher at higher temperatures deeper in the atmosphere, and once a parcel of air from this region containing that ratio convects to the upper troposphere, the conversion to a lower ortho-para H$_2$ fraction takes years due to colder temperatures higher up. Therefore, we can use the ortho-para H$_2$ ratio at altitude as a measure of vertical motion \citep{banfield_2005}.

\subsection{Question 6: Effects of Extreme Seasonal Forcing}\label{sec:2.7}

\subsubsection{Background and Motivation}\label{sec:2.6.1}

Neptune's and Uranus's circulation seem to be broadly similar despite Uranus's 98\textdegree obliquity and much higher degree of insolation at its poles relative to its equator. Uranus's deep atmosphere has displayed behavior consistent with seasonal changes after the southern solstice in 1985 \citep{hofstatder_2003}, and more recent Very Large Array observations have shown potential seasonal changes in the brightness temperature of the polar cyclone at Uranus's north pole \citep{akins_2023}. Bright and short-lived cloud features have been occasionally observed at the top of the atmosphere (e.g. \citet{hammel_2005}), and while it is yet unclear what causes these events and if those effects are seasonal in nature, they might be linked to cumulus convective outbreaks or local condensation resulting from vortices causing vertical displacement in the atmosphere \citep{sromovsky_2012}. In general, cloud and apparent convective activity increased as Uranus approached and passed the 2007 equinox, until 2014. After 2014, observable convective activity appeared to cease, with the exception of occasional white clouds near the polar hood \citep{hueso_2020}.

Remote-sensing instruments onboard an orbiter are likely best-suited for addressing this question, in that they can obtain observations over much more expansive temporal, spatial, and longitudinal ranges than a probe. However, a properly-equipped \textit{in situ} probe would be able to greatly enhance our ability to answer this question by directly probing the atmospheric structure and processes deeper down. Particularly, measuring the lapse rate and how it changes with altitude offers insight into the degree of vertical mixing and convective strength as a function of depth. How the upper and lower regions of Uranus's atmosphere influence one another is essential to understanding the planet's energy budget and the interplay between seasonal and non-seasonal forcings.

Specifically, several questions remain on the seasonal dynamics of the Uranian atmosphere. (1) How does the global circulation change on the seasonal timescale? Ground-based observations have shown that cloud and storm formation is tied to specific seasons \citep{dePater2015}, but is this a result of global upwelling and downwelling or localized orographic effects from polar vortices? (2) How does the thermal flux change on the seasonal timescale? We require more measurements of the thermal flux on Uranus over multiple seasons in order to understand this variability. (3) How do aerosol properties change seasonally? Do the haze particles vary significantly with seasonal forcing? It is vital that we answer these questions so as to generate a model that unifies the various processes that drive the atmosphere (i.e., dynamics, microphysics, thermal and mechanical forcing, etc.).

\subsubsection{Measurement Requirements}

A probe capable of measuring changes in local and planet-wide circulation, especially when compared to Voyager 2 measurements, would dramatically increase the science return of a mission seeking to address this question. As the probe descends through the atmosphere, measurements of the vertical and horizontal wind profiles will provide vastly improved constraints for 3D circulation models of Uranus's atmosphere. 

Global circulation modeling studies would greatly benefit from even a single \textit{in situ} data point, specifically in constraining the vertical profile of temperature, wind speeds and aerosol density, all to accuracy levels previously described. These properties will prove critical to constraining current dynamical models \citep[e.g.,][]{hammel_2009, helled_2010,Kaspi_et_al_2013,LeBeau2020}, particularly with relation to the processes that drive the deeper atmospheres.

Additionally, in order to holistically determine atmospheric composition, we require measurements of the scattering cross section and spectral identification of haze aerosol species in the upper atmosphere as much as the probe is able below the lowest pressure limit of 100 mbar, particularly as they pertain to seasonal concentrations of aerosols and hydrocarbons. The ratio of trace species to the bulk abundances (e.g., of C, N, O and S) enable estimates on the vertical mixing length and timescale, as well as the degree of vertical circulation.

Furthermore, arriving at Uranus before or during the 2049 equinox will provide the best opportunity to measure the atmosphere when the planet is at the largest disparity between measurement epochs by dedicated spacecraft. Voyager 2 arrived at Uranus during solstice and took relatively limited measurements due to the nature of its quick flyby; orbiting and probing Uranus during equinox will enable a direct comparison of the two spacecrafts' datasets at opposite seasonal epochs. Arriving during a time frame leading up to and/or after an equinox enables a comparative planetology experiment between Uranus and other bodies that are particular active at equinox, such as Titan \citep{Coustenis2020}.

\section{Optimized Payload for Atmospheric Science}\label{sec3}

A mass spectrometer will be critical for addressing the majority of our proposed science questions by generating vertical profiles of various atmospheric species. In particular, to address Question 1, a mass spectrometer is necessary to measure abundances of the noble gases and their isotopic ratios in order to utilize them as a point of comparison to the other giant planets, and to cometary and Solar abundances \citep{Vorburger_2020, cavalie_2020}. The mass spectrometer will also measure the abundances and ratios of C, N, S, O, and P through tropospheric measurements of CH$_4$, H$_2$S (or NH$_3$ if available to the probe), CO, and PH$_3$. 
During the probe's descent, the mass spectrometer should make a minimum of one measurement of noble gas and their isotope abundances at any altitude below the homopause to address Question 1 alone; additional measurements for redundancy would be helpful to provide further constraints on those data. For the other gas abundances, a minimum of one measurement every scale height would suffice to obtain their bulk abundances \citep{orton_2020_1, mousis_2018}. Detailed vertical profiles of condensable species abundances are not necessary to address Question 1, but a much finer resolution is required for meeting the measurement requirements of Questions 2, 3, and 4.

In addition to the mass spectrometer, a tunable laser spectrometer (TLS) \citep{durry_2002} would be useful for high-accuracy measurements of targeted isotopic species, enabling accurate measurements of isotopic ratios such as D/H, $^{13}$C/$^{12}$C, and $^{18}$O/$^{17}$O/$^{16}$O, reinforcing and complementing some of the mass spectrometer measurements \citep{Vorburger_2020}. Additionally, a He abundance detector capable of measuring the He/H$_2$ ratio would also be highly desirable and would enable direct comparisons to Galileo probe measurements in Jupiter's atmosphere \citep{von_zahn_1998} by acquiring a more accurate measurement of the He abundance than the mass spectrometer alone. Such an accuracy will be necessary to look for variations from Voyager observations \citep{conrath_1987} and differentiate scenarios in which Uranus's He abundance becomes sub-protosolar, protosolar, or super-protosolar \citep{guillot_2005, mousis_2018}. The TLS and the HAD would especially enhance our ability to address Questions 1 and 3.

An Atmospheric Structure Instrument (ASI) is required to accurately measure the altitude profile of atmospheric pressure, temperature, and electrical properties and contextualize mass spectrometer measurements \citep{ferri_2020}. Additionally, electrical conductivity sensors, included as a part of the ASI, will provide a clues about atmospheric electrification processes, which generate information about possible lightning generation and cloud formation as a function of depth (e.g. \citet{hamelin_2007}). This instrument will be critical for addressing Questions 2-5.

An Acoustic Anemometer (e.g., \citet{banfield_2005}) is capable of producing the ortho-para H$_2$ ratio, since those two forms of H have different thermodynamic properties and affect the speed of sound differently; in essence, an acoustic anemometer exploits the sonic speed differences between these forms of hydrogen molecules. The H$_2$ ortho-para fraction, which serves as an important dynamical tracer for vertical motion, is also a necessary context for compositional measurements. Relative to ortho-hydrogen, low levels of para-hydrogen suggest upwelling from the deeper atmosphere, whereas high levels of para-hydrogen suggest downwelling. See Section 2.3 of \citet{fletcher_2020} for more details of the implications of ortho-para hydrogen abundances. At each measurement, the local temperature and mean molecular weight must also be accounted for to accurately derive the ortho-para H$_2$ fraction from the speed of sound. The accuracy of the acoustic anemometer needs to be at $\pm$1\% of the sonic speed, as stated in \citet{orton_2020_1}. An acoustic anemometer will contribute to answering Questions 2, 3, and 5.

Questions 4 and 6, which both require the characterization of aerosols and any possibly measurable photochemical hazes, necessitate a nephelometer. The nephelometer onboard the Galileo probe would be a prime legacy instrument to use as a model \citep{ragent_1998}.

The vertical and (to a lesser degree) horizontal winds can be obtained through a Doppler wind experiment (DWE) \citep{atkinson_1998}. Accelerometers, either as a part of this instrument package, or as part of the engineering telemetry, can be inverted to produce a vertical profile of winds and measurement of atmospheric waves that may be encountered. It should be noted that horizontal wind measurements, which can be extracted from the horizontal winds' centrifugal force effects on measurements of vertical acceleration \citep{seiff_1997}, will have a low resolution due to noise from the turbulence encountered by the probe as it spins and falls. Temperature/pressure profiles from the ASI will also be necessary to contextualize the vertical wind profiles. The DWE and resulting measurements will help answer Questions 2-5.

A net flux radiometer (NFR) is an instrument made to measure the net atmospheric radiative energy balance by capturing the upward thermal infrared and downward solar radiative fluxes. To accurately measure the amount of heat flux from sunlight, the NFR must operate on the day side of the planet. It is capable of measurements from 0.1 to at least 10 bars. Depending on the probe's entry point, the NFR can provide valuable insight about regions of solar energy deposition and additional constraints on atmospheric composition and cloud properties  \citep{sromovksy_1998}. Measurements at least every scale height would be best, but more is better if data volume allows. This instrument is necessary for addressing Question 2.

To enhance the results of addressing these science questions, there are some complementary instruments that should and in some cases need to be included onboard the orbiter. 
The abundances of microwave absorbers, such as PH$_3$, and their vertical profiles can also be derived from radio science experiment measurements as the probe descends through the atmosphere \citep{howard_1992}. While not a probe instrument, a microwave radiometer onboard the orbiter can detect microwave absorbers below the pressure-level limit of the probe; the microwave radiometer onboard the Juno spacecraft could serve as an excellent legacy instrument \citep{janssen_2017}. Both of these measurements will provide important context for the abundances measured \textit{in situ} for Question 1.

In addition to these instruments onboard the atmospheric probe, a photometer or calibrated imager onboard the orbiter would be critical to measure the Bond albedo of the planet in order to account for the amount of reflected and absorbed sunlight, which is necessary for answering Question 2. This measurement would be used to account for the heat energy delivered to the planet by solar radiation \citep{pre_decadal_survey}. Along with measurements of Bond albedo, full phase angle coverage of the emitted thermal radiation from orbit is also vital to accurately quantify the amount of released heat from within the planet. This can be accomplished with a thermal IR bolometer on board the orbiter \citep{pre_decadal_survey}. 
To better illuminate the effects of seasonal forcing in support of Question 6, a long-period orbiter equipped with an optical or near-IR multi-spectral imager would also enable measurements of variations in aerosol opacity over long timescales and identify any seasonal changes in aerosol distribution, which would provide an excellent complementary dataset to probe measurements.

To best address Question 6, directly probing atmospheric circulation and the degree of vertical mixing at \textit{multiple} locations would be ideal. To that end, a spacecraft equipped with multiple probes, which might be dropped at multiple latitudes, in regions with different degrees of upwelling or downwelling, or just at different points in the mission might also help illuminate any possible effects of seasonal forcing \citep{sayanagi_2020}.

\section{Summary and Conclusions}\label{sec4}

As a result of several factors, including a great deal of community support, the potential scientific richness of a mission to Uranus, and simple technological readiness, NASEM's 2023-2032 Oceans, Worlds, and Life: A Decadal Strategy for Planetary Science and Astrobiology recommended a UOP flagship mission to be NASA's highest priority for development over the next decade \citep{owl}. An orbiter alone would open the door to an enigmatic, icy world, allowing for detailed remote-sensing measurements and the temporal coverage required to understand the variability and behavior of various aspects of the Uranian system. However, the addition of a probe, and its ability to obtain \textit{in situ} measurements, will enhance the science return of the mission tenfold and enable a degree of exploration unavailable to an orbiter alone. Within the new context of NASEM's recommendations, we here reassessed several community-sourced science questions originally developed for both Ice Giants and independent of the type of spacecraft that might fly. These questions were explored in terms of how an atmospheric probe might address them, through measurement and instrument requirements and suggestions. 

\subsection{Summary of Instrument Recommendations}

In summary, we recommend the following instrumental suite (in no particular order) for a Uranian probe in order to address the posed atmospheric science questions:

\begin{itemize}
    \item Atmospheric structure instrument and associated accelerometers to provide the capability for constructing pressure, temperature, and thermospheric density profiles as a function of pressure and depth
    \item Acoustic anemometer to determine the speed of sound, which will provide the hydrogen ortho-para fraction either as a probe of thermal properties or as a dynamical tracer
    \item Mass Spectrometer to measure noble gases, their isotopic ratio, and elemental/molecular abundances; maybe a tunable laser spectrometer and/or helium abundance detector to improve the accuracies of certain measurements
    \item Nephelometer to characterize cloud structure and aerosol properties
    \item Net flux radiometer to measure atmospheric radiative energy structure
    \item Doppler wind experiment to construct the vertical profile of horizontal winds
\end{itemize}

Additionally, in the case where certain instruments onboard the orbiter might complement or be required for fully contextualizing probe measurements, we recommend at minimum these remote-sensing instruments: 
\begin{itemize}
    \item Bolometer and/or imager to assess the level of reflected and absorbed sunlight for purposes of isolating the amount of released internal heat
    \item Thermal IR bolometer for measuring Uranus's total emitted thermal radiation
    \item A microwave radiometer for measurements of microwave-absorbing condensable species, such as NH$_3$, PH$_3$, and H$_2$S, which can be utilized to explore pressures too deep for the probe to access
    \item Multi-spectral imager (particularly in the visible and IR wavelength regimes) optimized to characterize aerosol property changes over time, in order to link them to any seasonal effects
\end{itemize}

\subsection{Additional Requirements}

Additional requirements for this atmospheric, specialized probe includes a depth range of at least 0.1-10 bars. Above $\sim$0.01 bars, the probe's parachute will not have deployed, and so it cannot make measurements at lower pressures. The parachute will likely be fully unfurled, the foreshield jettisoned, and the probe stable enough to make accurate measurements by 0.1 bars. A depth of 10 bars is sufficient to identify the most important condensable species (i.e. CH$_4$ and H$_2$S; note that the available H$_2$S abundance will be limited by the condensation of the NH$_4$SH cloud at deeper levels) and measure the temperature/pressure profile, but expanding those values down to larger pressures would only improve the context for other measurements, including those made by the orbiter. 

For an insertion location, the equator might be ideal for enabling direct comparisons to Juno microwave radiometer measurements of Jupiter’s unexpected ammonia distribution, and higher latitudes would aid with better dynamical constraints. To measure noble gas abundances and isotopic ratios of key elements to answer the composition and formation history question, the probe could enter at any location on the planet. For measurements seeking to address the questions regarding the internal heat budget, thermal evolution, and thermal flux, the entry location should be a region that can and will be comprehensively targeted by remote-sensing observations over the time span of the mission, since the \textit{in situ} measurements are essential to provide a ground-truth for remote-sensing observations. For other questions, including those focusing on moist convection and zonal/meridional circulation, a location that has differing vertical structures for these quantities predicted by different theoretical models would be ideal. However, current theories of moist convection and atmospheric circulation at Uranus are not yet mature enough to clearly define such locations. One solution for this problem is to use a multi-probe approach, which would enable comparisons of vertical structure at multiple latitudes (e.g. if two probes entered a pair of cyclonic and anticyclonic shear zones or into opposite-season hemispheres) will enable overcoming the insufficiency of theoretical understanding. In general, when measurements of aerosol properties are being collected, at least one measurement per scale height is required, but multiple measurements per scale height, as much as data volume allows, is ideal.

Recent probe studies conducted with Uranus in mind generally converge on a similar suite of required instruments. However, as the experience of the Galileo probe revealed, adding at least one other probe would greatly enhance the \textit{in situ} science the UOP will be able to complete. The secondary probe could be small ($\sim$30 kg) and would make limited measurements to complement the primary probe. For a thorough discussion of the possible roles of and science return from multiple probes, see \citet{sayanagi_2020} and \citet{wong++this-ssrv-issue}. Depending on when the UOP arrives at the planet, such a secondary probe could be targeted to enter the atmosphere in a different hemisphere or to a different latitudinal region within the same hemisphere to optimize context for the orbiter and primary probe's measurements. 

Arriving at Uranus at equinox, around the late 2040's would also be ideal. Launching the spacecraft in the early 2030's will allow it to take advantage of a Jupiter gravitational assist, thereby maximizing the amount of spacecraft mass that can be dedicated to instrumentation and allowing for comparisons between the seasons measured by Voyager and the UOP, should also be a top priority \citep{hof_2019}.

\subsection{Conclusions}

Overall, there is currently a great deal of community support for the UOP, highlighted by the number of recently-developed white papers, meeting abstracts, and peer-reviewed works outlining the benefits and possible mechanics of such a mission (e.g., \citet{owl}; \citet{hof_2019}; \citet{mousis_2020_2}). Uranus occupies a unique parameter space; it is an extreme laboratory for atmospheric, magnetic, and space physics unique within our Solar System. Exploring not only the atmosphere with \textit{in situ} measurements but long-term remote-sensing observations over a variety of viewing geometries and at spectral and spatial resolutions wholly unobtainable from ground-based observatories, will revolutionize our understanding of Ice Giants, Gas Giants, the formation and evolution of our Solar System, and that of exoplanetary systems beyond. 

\backmatter

\bmhead{Acknowledgments}

 Emma Dahl’s research was supported by an appointment to the NASA Postdoctoral Program at the Jet Propulsion Laboratory, administered by Oak Ridge Associated Universities under contract with NASA, and along with Glenn Orton and Shawn Brueshaber, by funds from NASA distributed to the Jet Propulsion Laboratory, California Institute of Technology. Kunio Sayanagi's efforts are supported in part by NASA Solar System Workings program grant 80NSSC21K0166 and NASA Cassini Data Analysis Program grant 80NSSC19K0894. Ramanakumar Sankar was supported in part by NASA Solar System Workings grant 80NSSC22K0804. The authors gratefully acknowledge the anonymous reviewers and David Atkinson for their helpful comments which improved the quality of this manuscript. This work was carried out at the Jet Propulsion Laboratory, California Institute of Technology, under a contract with the National Aeronautics and Space Administration (80NM0018D0004). \copyright 2023. All rights reserved. 

\section*{Declarations}

\begin{itemize}
\item The authors declare no conflicts of interest related this work. 

\end{itemize}

\bigskip

\bibliography{sn-bibliography}

\end{document}